\documentstyle[12pt,times]{article}

\newcommand{\be}{\begin{equation}}
\newcommand{\ee}{\end{equation}}

\newcommand{\al}{\alpha}
\newcommand{\bary}{\begin{eqnarray}}
\def \ua {\uparrow}
\def \da {\downarrow}
\newcommand{\gav}{{g_A}}
\newcommand{\eary}{\end{eqnarray}}
\begin{document}
\begin{titlepage}
\title{Axial Vector Coupling Constant in \\
Chiral Colour Dielectric Model}
{\author{Sarira Sahu$^{1}$\\          
Theory Group, Physical Research Laboratory,\\
Navrangpura, Ahmedabad-380 009,India}}
\date{ }
\footnotetext[1]{email:sarira@prl.ernet.in\\
Present address: Institute For Plasma Research,
Bhat, Gandhinagar-382 428, India}
\maketitle
\thispagestyle{empty}
\begin{abstract}

\noindent 
The  
axial vector coupling constants of the $\beta$ decay processes of
neutron and  hyperon are calculated in SU(3) chiral colour dielectric model 
(CCDM). Using these axial coupling constants of neutron and hyperon,
in CCDM we calculate the integrals of the spin dependent structure functions
for proton and neutron.
Our result is similar to the results obtained by MIT bag and
Cloudy bag models. 
\end{abstract}
\end{titlepage}
%\vfill
\eject

\section{Introduction}

In terms of quark parton model (QPM) interpretation the EMC\cite{as} 
result leads to a negligible contribution of the quark spins to the 
proton spin. This is in clear contradiction with the explanation given by the  
naive quark picture, in which
the proton spin comes from the sum of the quark spins. 
Moreover it is argued that the strange quark sea is polarized opposite to
that of the proton\cite{el,cl}. So the 
spin carried by $u$ and $d$ quarks cancel with
that of the strange quark. Thus the total helicity carried by the quarks is
very small.
The EMC experiment has been complimented by measurements of
the neutron spin structure function in the E142 experiment at SLAC and 
by the SMC deutron experiment\cite{slac}. All these combined datas are throughly
analysed by different groups\cite{el,cl}.

The Bjorken-sum\cite{bj} rule relates the polarized structure 
function for proton 
and neutron with the axial vector coupling constant
$g_A$ and the vector coupling constant $g_V$ of the nucleon and is given 
(without QCD correction) as  
\be
\int_0^1 \Big ( {g_1^p(x)}-{g_1^n(x)}\Big ) dx = {1\over 6}
|\frac{g_A}{g_V}|.
\label{bsr}
\ee
Similarly the Ellis-Jaffe (EJ)\cite{ej}
sum rule for proton and neutron are given as
\be
\int_0^1 g_1^{p(n)}(x)dx = {1\over 12}{g_A\over g_V}
\Big ( +(-)1+{5\over 3}{{3F/D-1}\over {F/D+1}}\Big ),
\ee
where F and D are antisymmetric and symmetric weak SU(3) couplings measurable
in the beta decay of the neutron and hyperons in baryon octet. 
This is derived with the assumptions that SU(3) flavour symmetry 
of the baryon octet is exact and strange quarks in the nucleon 
are unpolarized. In the above equations the
vector and axial vector coupling constants are calculated from the 
low-momentum transfer limit. But the polarized structure functions are
calculated from the high momentum transfer limit. So basically the 
above sum rules relate both the low and high momentum transfer phenomena.
By using the current algebra, one can show that the integral of the 
nucleon structure function can be given by the matrix element
\be
 \int_0^1 g_1^N(x)dx={1\over 2}
<N{\ua}|{\bar\psi}(0)Q^2\gamma_z\gamma_5{\psi}(0)|N{\ua}>,
\ee
where $g_1^N(x)$ in terms of the spin dependent quark distribution 
function is given as
\be
g_1^N(x)={1\over 2}\sum_i Q_i^2\Big [ q_f^{\ua}(x)-q_f^{\da}(x)\Big ].
\ee
Here $q_f^{{\ua}({\da})}(x)$ is the number density for quarks of flavour
$f$ and charge $Q_f$ at momentum fraction $x$, and with spin parallel 
(anti-parallel) to that of the nucleon. 
So we can calculate the axial coupling constants for  
neutron and hyperon beta decay processes from  
any phenomenological model of hadron and relate that to 
the structure functions. This is the basis of our calculation in this 
paper. We have used the SU(3) chiral
colour dielectric model (CCDM)\cite{ss} to calculate the 
axial coupling constants for the process $n\rightarrow pe^-{\bar\nu_e}$ and 
the hyperon beta decay ($\Sigma^-\rightarrow ne^-{\bar\nu_e}$)
process and use the above sum rules to predict
the integral of the spin structure functions of the nucleon.

The chiral symmetry has long been known to be an important symmetry
of the strong interaction. Massless QCD is invariant under this
symmetry. In the context of $\sigma$-model, Gell-Mann and Levy\cite{gel}
showed that nucleon acquires mass via the spontaneous breaking of
chiral symmetry. Later on it was realised that this
symmetry has to be incorporated in all the phenomenological
models of hadron.
In MIT bag model\cite{tho}, the chiral 
symmetry is violated due to the fact that, the 
reflected quark from the bag surface does not flip its spin, hence the quark
with helicity +1 after reflection will flip its helicity to $-1$, thus violating
the chirality. On the other hand, the bag model with pion (Cloudy bag model)
conserve the chiral symmetry\cite{tho}. 
The change in the helicity due to reflection
is compensated by the emission of a $p$-wave pion of unit orbital angular 
momentum leaving the bare three quark system (this system might be nucleon or
delta). 
So the quarks spin inside the baryon may not add up to
$\pm{1\over 2}$. So the spin structure of the nucleon reveled by the 
electromagnetic probe might be different.
The chiral SU(2) version of Cloudy bag model\cite{tho}, 
Friedberg-Lee 
soliton model\cite{de} and Colour Dielectric model have been studied
extensively\cite{mkb}. All these calculations yield a good 
agreement with the static hadronic properties. 
These models have been extended to include SU(3) chiral symmetry to study the 
pseudo scalar meson octet contribution\cite{ss,jo} although this symmetry is
badly broken due to the different masses of the pion, kaon and eta mesons.

\section{Model}

Unlike other models, the colour dielectric model generates the
absolute confinement dynamically.
The scalar field in the model lagrangian
takes into account the long range order effect due to
non-perturbative QCD vacuum and the short wave-length components do
not exist. As the scalar field takes into account the long
distance behaviour, the gluonic fields are treated perturbatively.
For $\chi \rightarrow$ 0, the inverse coupling of scalar field to
quarks i.e. ${m_q\over \chi}{\bar \psi}\psi$ dynamically confines the
quarks; even in the absence of gauge field. The chiral SU(2) version
of CDM has been successfully used to study the nucleon static
properties. The SU(3) chiral colour dielectric model has also been used to
study the strangeness related phenomena. 

The SU(3) CCDM lagrangian is given 
by\cite{ss}
\bary
{\cal L} &=& \sum_i{\bar \psi_i} \Big [ i\gamma^{\mu}\partial_{\mu}
-\Big ( m_{su}+{m_u\over \chi}(1+{i\over
f_{\phi}}\gamma_5\lambda.\phi)\Big ) 
-{1\over 2} g_s\gamma^{\mu}\lambda_a A^a_{\mu}\Big ]\psi_i
\nonumber\\
& &-{1\over 4} \kappa(\chi)F^a_{\mu\nu}F^{a,\mu\nu}
+{1\over 2}\sigma^2_v({\partial_{\mu}\chi})^2
+{1\over 2}(\partial_{\mu}{\vec\phi})^2-{1\over 2}m^2_{\phi}{\phi}^2-U(\chi),
\eary
where $\psi$, $\chi$, $\phi$ and $A^a_{\mu}$ are effective quark,
colour dielectric, SU(3) pseudo scalar meson and the gluon fields respectively.
$m_{\phi}$ is the octet meson mass, $f_{\phi}$ is
the meson decay constant, $\alpha_s = {g_s^2}/{4\pi}$ is the
strong coupling constant and $F^{a,\mu\nu}$ is the colour electromagnetic 
field tensor. The 
sum $i$ is over quark colour and flavour and the effective quark mass 
is 
($m_{su}+{m_u\over \chi}$) with $m_{su}=0$ for $u$ and $d$ quarks.  
The gluon field
interacts with the dielectric field through a dielectric functional
$\kappa (\chi) = \chi^4 (x)$ and the self-interaction of the dielectric
field is given as 
\be
U(\chi) = B \big [\alpha \chi^2-2(\alpha-2)\chi^3+(\alpha-3)\chi^4
\big ].
\ee
For $\alpha > 6$,  $U(\chi)$ has a double well structure, with an
absolute minimum at $\chi$=0 and local minimum at $\chi$=1 and the
energy density difference between these two minima gives the bag
constant B. The mass of the scalar field is given as 
$m_{GB}= \sqrt{2B\alpha\over \sigma_v^2}$ and this is interpreted as the
glueball mass. The strong coupling constant is calculated  
by fitting the nucleon and the delta masses.

In SU(3) CCDM picture the physical baryon is a system of three quarks
surrounded by a meson cloud.
   Following the approach of Thomas et al.\cite{tho}, 
   the physical baryon state $\mid A>$ can be expressed as
\be
\mid A> ={\sqrt{P_A}}\Big \{ 1+(m_A-{\tilde H_0})^{-1}H_{int}\Big
\} \mid A_0>.
\ee
Here $P_A$ is the probability of finding bare baryon (three quarks) state
$\mid A_0>$ with bare mass $m_{A0}$ and $m_A$ is the physical baryon mass. 
$H_0$ is the noninteracting hamiltonian which includes quark and the
dielectric field hamiltonian and free meson part. $H_{int}$ is the quark-
meson interaction hamiltonian. The second term in the above equation is 
responsible for the generation of meson cloud around the bare baryon.
In this picture the meson field is considered to be small, so that non-
linearities  due to meson coupling can be neglected and meson contribution
can be treated perturbatively.

\section{Semileptonic Decay}

The $\beta$ decay process between octet baryons $\bf A$ and 
B is
$A\rightarrow Be^-{\bar \nu}_e$.
For the calculation of the axial coupling constant (semileptonic
decay constant) we
have to evaluate the matrix element of the axial vector current (isospin
changing)
$
<B\mid A_{\mu}(x)\mid A>
$,
where $\mid A>$ and $\mid B>$ are initial and final baryon states,
and the axial vector current is 
\begin{eqnarray}
A_{\mu}(x) &=& {\bar \psi}(x) \gamma_{\mu} \gamma_5
{\lambda\over 2}\psi(x)
\nonumber\\
&=& {\bar \psi}(x) \gamma_{\mu} \gamma_5
{1\over 2} 
\big({\lambda_1+i\lambda_2}\big)\psi(x)
+{\bar \psi}(x) \gamma_{\mu} \gamma_5
{1\over 2} 
\big({\lambda_4+i\lambda_5}\big)\psi(x)
\nonumber\\
&=& A_{\mu 1}(x)+A_{\mu 2}(x).
\label{axl}
\end{eqnarray}
In the Eq.(\ref{axl}) the first term of the RHS correspond to 
strangeness change 0 ($\Delta S~=~0$) and the second term  
is for strangeness change 1 ($\Delta S~=~1$).

Since the quark-meson coupling is linear in the meson field and the 
mesonic part of the axial current is proportional to the divergence
of the the meson field, there will be no contribution to semileptonic
decay constant from the meson cloud\cite{ta}.
Thus the contribution to
the semileptonic decay constant comes from the quark part of the
axial current only.
The quark part of the axial coupling constant is calculated from
the matrix element\cite{ss,tho}
$
<B\mid A_{\mu}(x)\mid A>
$ and is given as
\bary
g_{A} &=&
<B\mid A_{z}(x)\mid A>
\nonumber\\
&=&
{\sqrt {P_A P_B}} <B_0\mid A_z(x)\mid A_0>
\nonumber\\
& & + {\sqrt {P_A P_B}}
<B_0\mid H_{int} (m_B-{\tilde H}_0)^{-1}
{A_z(x)} (m_A-{\tilde H}_0)^{-1}H_{int}\mid A_0>.
\eary
Summing over the intermediate states and after some algebra,
the second term in the above equation can be
expressed in terms of $3j$ symbols and can be written as
\bary
\lefteqn{
{\sqrt {P_A P_B}}
<B_0\mid H_{int} (m_B-{\tilde H}_0)^{-1}
{A_z(x)} (m_A-{\tilde H}_0)^{-1}H_{int}\mid A_0>} & &
\nonumber\\
&=& {{\sqrt {P_A P_B}}\over {3\pi}}
\sum_{C,D}
\Big ( {f_{BC\phi}f_{AD\phi}\over m^2_{\phi}} \Big )
2\eta {\sqrt {(2T_A+1)(2T_B+1)}}
\nonumber\\
& &\times
\sum_{m_1,m_2,i}
{\pmatrix {{1\over 2} & 1 & S_C\cr -s_B & m_1 & s_C\cr}}
{\pmatrix {{1\over 2} & 1 & S_D\cr -s_A & m_2 & s_D\cr}}\nonumber\\
& &\times
{\pmatrix {T_B & T & T_C\cr -t_B & i & t_C\cr}}
{\pmatrix {T_A & T & T_D\cr -t_A & i & t_D\cr}}
\nonumber\\
& &\times
\int dk {{k^4 u_{BC}(k)u_{AD}(k) <C_0\mid {A_z}\mid D_0>}
\over {(\omega_{CB}+\omega_k)(\omega_{DA}+\omega_k)\omega_k}}
\nonumber\\
&=& {\sqrt {P_A P_B}}
\sum_{C,D} g(C,D),
\eary
where $f_{BC\phi}$ and $f_{AD\phi}$ are the baryon-meson coupling
constants, $u(k)$ is the baryon-meson form factor and
$\omega_k={\sqrt{(k^2+m_{\phi}^2)}}$.
The phase factor is
\be
\eta=(-1)^{(T_A+T_B-s_A-s_B-t_A-t_B+1)}.
\ee
Thus the axial coupling constant is given as 
\be
g_{A} = 
{\sqrt {P_A P_B}}
\Big [ <B_0\mid {A_z(x)}\mid A_0>
+\sum_{C,D} g (C,D)\Big ].
\label{interm}
\ee
The matrix elements $<B_0\mid {A_z}(x)\mid A_0>$ 
and $<C_0\mid {A_z}(x)\mid D_0>$ can be evaluated using the wave functions
of the respective baryons. The second term in Eq.(\ref{interm}) corresponds
to the meson exchange contribution.

\section{Results}

Briefly, the calculation in perturbative CCDM proceeds 
as follows. One first 
solves the quark and dielectric field equations in mean field 
approximation and constructs the (bare) baryon states consisting of 
three quarks. In our calculation Peierls-Yoccoz momentum 
projection technique\cite{mkb,dd} is used to construct good momentum 
states and these
states are used to calculate the bare baryon properties
and baryon-meson
form factors. The pseudo scalar 
meson-quark interaction is then included to calculate the mesonic 
effects on baryon properties. 

The parameters in the CCDM are $m_{GB}$, $m_u$, $m_{su}$, $\al_s$, 
B, $f_\phi$ and $\al$. Our calculations show that the parameter $\al$, 
which determines the height of the maximum of $U({\chi})$ between 
two minima at $\chi=0$ and $\chi=1$ does not play an important role 
in the calculation. We have chosen $\al=36$ throughout. Also, 
the meson-quark 
coupling constant $f_\phi$ has been chosen to be 93 MeV, the pion decay 
constant. The rest of the parameters are varied to fit the properties 
of octet and decuplet baryons. We find that the masses of these 
baryons can be fitted, to a very good accuracy, for a family of parameter 
sets. 
The numerical results for our calculation for  different parameter
sets are shown in table 1. 
For all these cases, the baryon octet and
decuplet masses are reproduced very well. 
There is, however,
a large variation in the calculated static properties (charge radius,
magnetic moment, axial coupling constant etc.). Generally small
glueball masses give a reasonable agreement with the static properties
(except for magnetic
moments). As the glueball mass is increased, magnitudes of charge radii,
magnetic moments and axial
coupling constant decrease. This can be attributed  to the increase
in percentage of meson cloud with the increase in the glueball mass\cite{ss}.
We have also shown the results obtained 
from different quark models\cite{th2}.
We have shown results for four parameter sets. The four parameter sets
are given as follows:

\noindent
Set A: $m_{GB}$=1050 MeV, $m_u$=105 MeV, $m_s$=318 MeV and $B^{1/4}$=94.5 MeV\\
Set B: $m_{GB}$=804 MeV, $m_u$=88 MeV, $m_s$=307 MeV and $B^{1/4}$=88.4 MeV\\
Set C: $m_{GB}$=4019 MeV, $m_u$=80 MeV, $m_s$=294 MeV and  $B^{1/4}$=180 MeV\\
Set D: $m_{GB}$=1016 MeV, $m_u$=38 MeV, $m_s$=311 MeV and $B^{1/4}$=114 MeV\\

The table 1 shows the $\gav$
for neutron and $\Sigma$ 
beta decays ($\gav_{np}$ and 
$\gav_{\Sigma n}$). 
The $g_V$ of nucleon and hyperons are normalized to unity.
We have used these two coupling constants to calculate the F/D ratio.
Our calculation shows that the $\gav$ for above two decay processes
are less than the observed values. 
This is because the bare probabilities
for nucleon $(P_n)$ and sigma $(P_{\Sigma})$ are less than one. 
For example for the parameters of row one of the table, the probability 
of bare nucleon and sigma are respective 0.83 and 0.87. So the inclusion
of meson  reduces the $\gav$ for neutron and $\Sigma$ by 17\% and 13\%
respectively.
This reduction could have compensated
by meson exchange contribution. But it is observed that the meson 
contribution is very small to over come the reduction. Also the 
kaon and eta contributions are opposite to 
that of the pion contribution.
So this also reduces the axial coupling constant. 

The F and D calculated from the above two baryon decays are shown in the
table. It shows that for large glueball mass (Set C) these two SU(3)
couplings are small compared to the one obtained from comparatively
smaller glueball masses. But the ratio F/D is constant through out. 
We found the ratio F/D$\simeq$0.67 over wide range variation of parameters.
This ratio is higher than the 
value obtained by Close and Roberts\cite{cl} and 
Ehrnsperger et al.\cite{es}
The values for F and 
D obtained by Close et al. are 
$0.459\pm0.008$ and $0.798\pm0.008$
(F/D=$0.575\pm0.016$)
respectively and the value for F/D in ref[\cite{es}] is 0.49$\pm$0.08.

The calculation of EJ sum rule for proton shows that our results
agree with the old result\cite{ej} and very large compared to the EMC
measurement\cite{as}. 
Comparison with MIT bag model and CBM\cite{th2}, 
shows that all the results are similar. 
We get small, non-zero and negative value for the 
neutron spin structure function, which is consistent with the 
experimental data.
This non-zero contribution is solely
attributed due to the meson exchange term. But this value is too small
compared to the observed  one.
On the other hand neutron structure function is zero in 
non relativistic quark model (NRQM) and MIT 
bag model as shown in the
table\cite{th2}. The NRQM gives a very 
large value of the $\gav$ for neutron beta decay.
Also it is seen that the proton structure function obtained using
this coupling constant is very high. 

Finally, our calculation shows that the proton spin structure function
calculated using the axial coupling constant 
is very large compared to the EMC result. It agrees with
the old result and  also with MIT bag and CBM results. 
On the other hand the meson cloud in  CCDM and CBM give a non-zero
and negative contribution to the neutron spin structure function,
which is consistent with the analysis of the recent experimental
datas. The above analysis of our results show that the
CCDM overestimate the value of the proton spin structure
function. Infact it is true for all the quark models.

\noindent{\bf Acknowledgment}

I would like to thank Prof. E. Oset, Prof. V. Vento, Prof. A. Pich and
Prof. Ulf-G. Meissner for many helpful
discussions. 
\vfill
\eject
\newpage
\begin{table}
\caption{
The results for four parameter sets of CCDM 
are shown in the table. The NRQM, MIT bag and
CBM results  are also shown from ref[14]. 
The $\gav_{np}$ for CBM is normalized to the 
observed value at 0.8 fm of the proton charge radius. 
The experimental values for $\gav_{np}$
=1.254$\pm$.006, 
$\gav_{\Sigma n}$=0.340, $\int g^p_1(x) dx$=0.114$\pm$0.012$\pm$0.026 and 
$\int g_1^n(x) dx$=-0.077$\pm$0.012$\pm$0.026.
The value of F=0.459$\pm$0.008, D=0.798$\pm$0.008 and
F/D=0.575$\pm$.016 in ref[4] and F/D=0.49$\pm$0.08 in ref[16].}
\vspace {0.5in}
\begin{center}
\begin{tabular}{|c|c|c|c|c|c|c|}
\hline
\multicolumn{1}{|c|}{$~~$}&
\multicolumn{1}{|c|}{$\gav_{np}$}&
\multicolumn{1}{|c|}{$\gav_{\Sigma n}$}&
\multicolumn{1}{|c|}{$\int g_1^p(x) dx$}&
\multicolumn{1}{|c|}{$\int g_1^n(x) dx$}&
\multicolumn{1}{|c|}{$F$}&
\multicolumn{1}{|c|}{$D$}\\
\hline
\hline
A  & 1.141 & 0.229 & 0.190 & -0.00032 & 0.456 & 0.685\\
\hline
B  & 1.166 & 0.234 & 0.194 & -0.00026 & 0.466 & 0.700\\
\hline
C  & 0.908 & 0.185 & 0.150 & -0.00095 & 0.361 & 0.546\\
\hline
D  & 1.180 & 0.237 & 0.197 & -0.00018 & 0.472 & 0.709\\
\hline
NRQM & 1.67 &      & 0.28  &  0       &       &      \\
\hline
MIT bag & 1.09 &    & 0.18 & 0        &       &       \\
\hline
CBM  & 1.254   &    & 0.193 & -0.010  &       &        \\
\hline
\end{tabular}
\end{center}
\end{table}
\vspace*{3cm}
\vfill
\eject


\newpage

\begin{thebibliography}{99}
\bibitem{as} European Muon Collab., J. Ashman et al., {\it Phys. Lett.} 
{\bf B206}, 364 (1988).
\bibitem{el} J. Ellis and R. A. Flores, CERN preprint \# TH-4911/87 (1987);
S. J. Brodsky, J. Ellis and M. Karliner, {\it Phys. Lett.}
{\bf B206}, 309 (1988);
J. Ellis, {\it Nucl. Phys.} {\bf A546}, 447c (1992);
F. E. Close and R. G. Roberts, 
{\it Phys. Rev. Lett.} {\bf 60}, 1471 
(1988);
\bibitem{cl} F. E. Close and R. G. Roberts, 
{\it Phys. Lett.} {\bf B316}, 165 (1993). 
\bibitem{slac} E142 Collab., P. L. Anthony et al., {\it Phys. Rev. Lett.}
{\bf 71}, 959 (1993); 
SMC Collab., B. Adeva et al., {\it Phys. Lett.} {\bf B302}, 533 (1993).
\bibitem{bj} J. Bjorken, {\it Phys. Rev.} {\bf 148}, 1467 (1966); 
{\bf D1}, 1376 (1970).
\bibitem{ej} J. Ellis and R. Jaffe, {\it Phys. Rev}
{\bf D9}, 1444 (1974); {\bf D10}, 1669 (1974).
\bibitem{ss} J. A. McGovern, {\it Nucl. Phys.} {\bf A533}, 533 (1991);
S. Sahu, S. C. Phatak, 
{\it Mod. Phys. Lett.} {\bf A7}, 709 (1992);
 S. Sahu, Ph. D thesis, Utkal University,
1993 (unpublished);
S. C. Phatak and S. Sahu, {\it Phys. Lett.} 
{\bf B255}, 11 (1994).
\bibitem{gel} M. Gell-Mann and M. Levy, {\it Nuovo Cim.} {\bf 16}, 705 (1960).
\bibitem{tho} A. W. Thomas, {\it Adv. Nucl. Phys.} {\bf 13}, 1 (1984);
A. W. Thomas, S. Theberge and G. A. Miller, {\it Phys.
Rev.} {\bf D24}, 216 (1981);
S. Theberge and A. W. Thomas, {\it Nucl. Phys.} {\bf A393},
252 (1983).
\bibitem{de} F. L. Dethier, Ph. D. thesis, Univ. of Washington, 1985;
F. L. Dethier and L. Wilets, {\it Phys. Rev.} {\bf D34}, 207 (1986);
M. C. Birse, in {\it Prog. Part. Nucl. Phys.}, ed. A. Faessler
(Pergamon, 1990), vol.25, p.1. 
\bibitem{mkb} 
M. K. Banerjee, {\it Phys. Rev.} {\bf C45}, 1359 (1992);
R. G. Leech and M. C. Birse, {\it Nucl. Phys.} {\bf A528}, 589 (1991); 
R. G. Leech and M. C. Birse, {\it J. Phys.} {\bf G18}, 589 (1992);
H. Kitagawa, {\it Nucl. Phys.} {\bf A487}, 544 (1988).
\bibitem{jo} J. A. Johnstone, {\it Phys. Rev.} {\bf D34}, 1499 (1986);
P. Gonzalez and V. Vento, {\it Nucl. Phys.} {\bf A407}, 349 (1983);
P. Zenczykowski, {\it Phys. Rev.} {\bf D29}, 577 (1984);
E. A. Veit et al., {\it Phys. Rev.} {\bf D31}, 1033 (1985).
\bibitem{dd} A. G. Willims and L. R. Dodd, 
{\it Phys. Rev.} {\bf D37}, 1971(1988);
H. J. Pirner, {\it Prog. Part. Nucl. Phys.} ed. A. Faessler
(Pergamon, 1992), vol29, p.33;
S. Sahu, {\it Nucl. Phys.} {\bf A 554}, 721 (1993);
S. Sahu, {\it Int. J. Mod. Phys.} {\bf E3}, 804 (1994);
M. Fiolhais et. al, {\it Nucl. Phys.} {\bf A481}, 727 (1988);
T. Neuber et. al, {\it Nucl. Phys.} {\bf A 560}, 909 (1993);
V. Barone and A. Drago, {\it Nucl. Phys.} {\bf A552}, 479 (1993);
H. Kitagawa, {\it Nucl. Phys.} {\bf A519}, 721 (1990).  
\bibitem{ta} R. Tang and W. Weise et al., {\it Phys. Lett.} {\bf B125}
9 (1983);
N. Barik and B. K. Das, {\it Phys. Rev.} {\bf D34}, 2092 (1986). 
\bibitem{th2} A. W. Sehreiber and A. W. Thomas, {\it Phys. Lett.}
{\bf B215}, 141 (1988).
\bibitem{es} B. Ehrnsperger and A. Sch\"afer, {\it Phys. Lett.} {\bf B348},
619 (1994).
\end{thebibliography}
\end{document}